\documentclass{PoS}

\title{$B_K$ with dynamical overlap fermions}

\ShortTitle{$B_K$ with dynamical overlap fermions}

\author{JLQCD Collaboration:
\speaker{N.~Yamada}\thanks{E-mail: norikazu.yamada@kek.jp}\ $^{a,b}$,~
S.~Aoki$^{\,c,d}$, H.~Fukaya$^{\,e}$,
S.~Hashimoto$^{a,b}$, J.~Noaki$^{a}$, T.~Kaneko$^{a,b}$,
H.~Matsufuru$^a$,  T.~Onogi$^f$\vspace{1ex}\\
\llap{$^a$}High Energy Accelerator Research Organization (KEK), Tsukuba
           305-0801,Japan\\
\llap{$^b$}School of High Energy Accelerator Science, The Graduate
           University for Advanced Studies (Sokendai), Tsukuba
           305-0801, Japan\\
\llap{$^c$}Graduate School of Pure and Applied Sciences, University of
           Tsukuba, Tsukuba 305-8571, Japan\\
\llap{$^d$}Riken BNL Research Center, Brookhaven National Laboratory,
           Upton, New York 11973, USA\\
\llap{$^e$}Theoretical Physics Laboratory, RIKEN, Wako 351-0198, Japan\\
\llap{$^f$}Yukawa Institute for Theoretical Physics, Kyoto University,
           Kyoto 606-8502, Japan
}
\abstract{We report on a calculation of $B_K$ with two-flavor
dynamical overlap fermions on a $16^3 \times 32$ lattice at $a\sim 0.12$
fm.
The results are compared with the PQChPT prediction of quark mass
dependence. The systematic errors due to finite volume effects and
fixing topology are discussed.}

\FullConference{The XXV International Symposium on Lattice Field Theory\\
		 July 30-4 August 2007\\
		 Regensburg, Germany}

\begin{document}

\section{Introduction}
\label{sec:introduction}

Indirect CP violation in $K$ decays, quantified by $|\epsilon|$,
has been playing an important role in finding the location of the apex of
the unitarity triangle in the $\rho$-$\eta$ plane and in constraining new
physics, especially the structure of flavor changing neutral current in
it.
Experimentally $|\epsilon|$ has been determined precisely as
$|\epsilon|$=(2.233$\pm$0.015)$\times 10^{-3}$~\cite{Yao:2006px}.
Within the standard model, $|\epsilon|$ can be expressed as
\begin{eqnarray}
 |\epsilon|=(\mbox{known factor})\times B_K(\mu) \times
           f(\bar\rho,\bar\eta),
\label{eq:epsilon}
\end{eqnarray}
where $f(\bar\rho,\bar\eta)$ is a known function of the Wolfenstein
parameters, $\bar\rho$ and $\bar\eta$, and
\begin{eqnarray}
   B_K(\mu)
 = \frac{\langle \overline{K^0} |\,
    \bar d \gamma_\mu(1-\gamma_5)s\
    \bar d \gamma_\mu(1-\gamma_5)s\,
   | K^0 \rangle}
   {\frac{8}{3}f_K^2 m_K^2}.
   \label{eq:bk definition}
\end{eqnarray}
%$\mu$ is the renormalization scale $\mu$.
The purpose of this work is to determine the parameter $B_K$ with high
precision using lattice QCD to give a strong constraints on $\bar\rho$
and $\bar\eta$ through eq.~(\ref{eq:epsilon}).

As seen from eq.~(\ref{eq:bk definition}), the $\Delta S=2$ four-quark
operator has the form of $(V-A)\times(V-A)$.
This makes the lattice calculation of $B_K$ much simpler if we
take the overlap fermion formalism because the overlap fermions respect
the lattice variant of chiral symmetry exactly at a finite lattice
spacing, and as a consequence the mixing with operators with other
chiralities is prohibited.
This simplification makes a precision lattice calculation possible.

%\section{Simulation parameters}
%\label{sec:parameters}

We perform the calculation on a $16^3\times 32$ lattice using the RG
Iwasaki action at $\beta=2.30$.
To accelerate HMC, we have introduced extra Wilson quarks and
ghosts~\cite{Fukaya:2006vs}.
At a price for the acceleration, the topological charge is frozen during
the HMC evolution.
Because of this, configurations are generated at a fixed topological
charge $Q=0$.
Six sea quark masses are taken in the range of [0.015,0.100] in lattice
unit, which roughly corresponds to
$[1/6\times m_s^{\rm phys}, m_s^{\rm phys}]$ in physical unit. 
Our lightest pion mass is about 290 MeV, and gives $m_\pi L\sim$2.7.
The lattice spacing $1/a=1.67(2)(2)$ GeV is determined by $r_0=0.49$ fm
in the $Q=0$ sector.
The physical spatial volume of our lattice is about (1.9 fm)$^3$.
In order to study the topological charge dependence, we have also
generated configurations at $Q=-2$ and $-4$ at $m_{\rm sea}$=0.050.
We have accumulated 10,000 trajectories for $Q=0$ and 5,000 for
$Q=-2$ and $-4$.

Calculations are done every 20 trajectories at each $m_{\rm sea}$.
Six valence quark masses take the same values as those of sea quarks.
All degenerate and non-degenerate mesons have been calculated.
We employ Coulomb gauge for the gauge fixing condition except for the
calculation of non-perturbative renormalization constant, in which
Landau gauge is used.
Low-mode averaging is implemented for all correlation functions,
which substantially improves statistical signals.

\section{Method and results}

Two-point functions are obtained in the standard way with a wall source
at $t_{\rm src}$ and a point sink at $t$.
We repeat this calculation four times with $t_{\rm src}$=0, 8, 16, 24,
and take an average over them.

The axial two-point function is defined by and fitted to
\begin{eqnarray}
     C^{(2),\rm p-w}_{A_4A_4}(t)
 &=& \sum_{\vec x}
     \langle 0\,|\,A_4(t,\vec x)\,
     (A_4^{\rm wall}(0))^{\dag}\,
     |\,0\rangle
 \rightarrow
     \frac{V_3\,Z^{\rm wall}_{A_4}}{2\,m_P}
     f_P\,m_P\,
     \left(e^{-m_P\,t}+e^{m_P\,(t-32)}
     \right),
     \label{eq:2-pt}
\end{eqnarray}
where
\begin{eqnarray}
 &&  A_4(t,\vec x)
 = \bar q_1(t,\vec x)\,\gamma_4\gamma_5\,q_2'(t,\vec x),\hspace{3ex}
     A^{\rm wall}_4(t)
  = \left(\sum_{\vec x} \bar q_1(t,\vec x)\right)\,\gamma_4\gamma_5\,
     \left(\sum_{\vec y}      q_2(t,\vec y)\right),\\
 && q_2'(x)=\left[1-D_{\rm ov}/(2\,m_0)\right]q_2(x),\ \
   Z^{\rm wall}_{A_4}
 = \langle P|\,\sum_{\vec x}
     \bar q_1(0,\vec x)\,\gamma_4\gamma_5\,q_2(0,\vec 0)\,
     |\,0\rangle,
\end{eqnarray}
$f_P\,m_P=\langle0|A_4|P\rangle$, $m_0=1.6$ and $V_3=16^3$.
$m_P$ and $Z^{\rm wall}_{A_4}f_P$ are extracted by a correlated fit.

In the calculation of three-point functions, the meson (anti-meson)
interpolating operators with wall source are put at fixed time slice
$t_1$ ($t_2$) while the position of four quark operator $t$ is varied.
%as shown in Fig.~\ref{fig:3pt-image-1}.
%\begin{figure}
% \centering
% \vspace{-0ex}
%  \includegraphics*[width=0.35 \textwidth,clip=true]
%  {figs/method0.eps}
% \caption{Setup for the calculation of three-point functions.}
% \label{fig:3pt-image-1}
% \vspace{-0ex}
%\end{figure}
Three-point functions are repeatedly calculated with
$(t_2,t_1)$=$(8,0)$, $(0,24)$, $(16,8)$, $(24,16)$, $(16,0)$, $(24,8)$.
These six pairs of $t_1$ and $t_2$ are classified into two sets by the
time separation, $|t_2-t_1|$=8 (24) or 16, which we call set A and B,
respectively.
Within each set, all the three-point functions are equivalent after
proper translation in the time direction, so they are averaged after
shifting.
As for set B (with $|t_2-t_1|$=16), two equivalent regions, $0<t<16$ and
$16<t<32$ are further averaged.
%Finally, we shift three-point functions to $(t_2,t_1)=(32
%(0),8)$ for set A and to $(t_2,t_1)=(16,0)$ for set B.

The three-point function is defined by eq.~(\ref{eq:3pt-2nd-def}) and
fitted to the form in eq.~(\ref{eq:3pt-2nd}):
\begin{eqnarray}
     C^{(3)}_{L_\mu L_\mu}(t_2,t,t_1)
 &=& \sum_{\vec x} \langle 0\,|\
     (A_4^{\rm wall}(t_2))^{\dag}\
     O^{\rm lat}_{L_\mu L_\mu}(t,\vec x)\
     (A^{\rm wall}_4(t_1))^{\dag}\
     |\,0\rangle
 \label{eq:3pt-2nd-def}\\
 &\rightarrow&
     \frac{V_3\,(Z^{\rm wall}_{A_4})^2}{(2\,m_P)^2}\,
     \langle \bar P|O^{\rm lat}_{L_\mu L_\mu}\ |P \rangle\,
     e^{-m_P(t_2-t_1)}
\nonumber\\
 &&+ \frac{V_3\,Z^{\rm wall}_{A_4}\,Z^{'\rm wall}_{A_4}}
         {2\,m_{P'}\,m_P}
    \langle {\bar P}'|\,O^{\rm lat}_{L_\mu L_\mu}\,|P\rangle\,
    e^{-(m_{P'}+m_P)\frac{t_2-t_1}{2}}
\nonumber\\&&\ \ \times
    \cosh\left[(m_{P'}-m_P)
               \bigg(t-\frac{t_2+t_1}{2}\bigg)\right]
\nonumber\\
 &&+ \frac{V_3\,(Z^{\rm wall}_{A_4})^2}{2\,(2\,m_P+\Delta_P)\,m_P}\,
     \langle 0|O^{\rm lat}_{L_\mu L_\mu}\ |P, P \rangle\,
     e^{-m_P\,N_t-\Delta_P (t_2-t_1)/2}
\nonumber\\&&\ \ \times
     \cosh\bigg[(2\,m_P+\Delta_P)
      \left(t-\frac{t_2+t_1}{2}\right)\bigg].
 \label{eq:3pt-2nd}
\end{eqnarray}
$O^{\rm lat}_{L_\mu L_\mu}
=\bar q_1 \gamma_\mu(1-\gamma_5)q_2'\
 \bar q_1\gamma_\mu(1-\gamma_5)q_2'$ is the $\Delta S$=2 four-quark
operator defined on the lattice, and $N_t=32$.
The first term in eq.~(\ref{eq:3pt-2nd}) contains the hadron matrix
element relevant to the calculation of $B_K$.
Since the time direction of our lattice is not so large, we add two
additional terms to represent an excited state contamination and a
contribution wrapping around the lattice.
The mass of the excited state $m_{P'}$ appearing in the second term
is extracted from the point-point pseudoscalar two-point function.
We confirmed that $m_{P'}$ are consistent with the experimental value of
$\pi(1300)$ in the chiral limit within the error.
%Notice that the excited state contamination of
%$\langle {\bar P}'|\,O^{\rm lat}_{L_\mu L_\mu}\,|P'\rangle$ is neglected
%as its effect to $B_P$ is estimated to be 0.03 \% at most.
The third term, expressing a wrapping contribution, contains a two-meson
system, and the energy shift $\Delta_P = E_{\rm total} - 2 m_P$
is extracted from the fit.

\begin{figure}
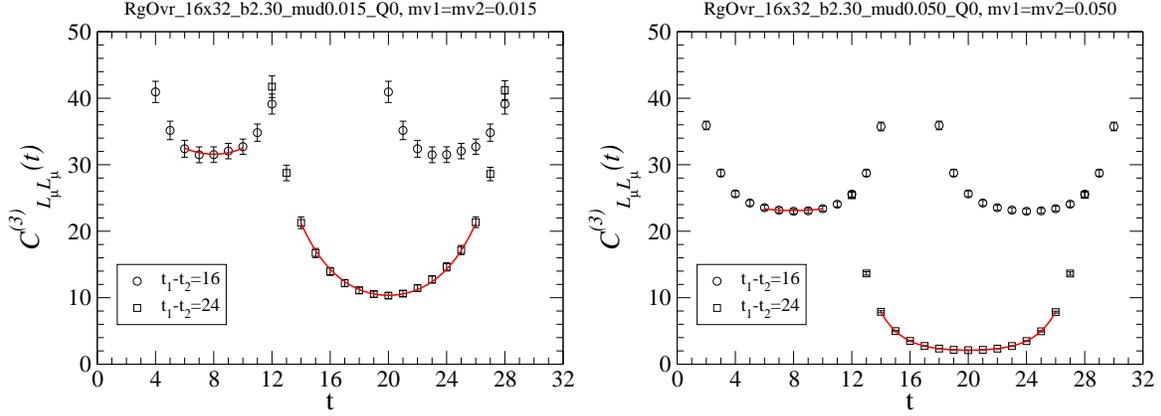

 \centering
 \vspace{-2ex}
  \includegraphics*[width=0.5 \textwidth,clip=true]
  {figs/3pt_LL_m0.015_001001.eps}~
  \includegraphics*[width=0.5 \textwidth,clip=true]
  {figs/3pt_LL_m0.050_004004.eps}
  \vspace{-4ex}
 \caption{Three-point functions. Data and fit results are shown.
 The fit range is $[t_{\rm min},t_{\rm max}]=[14,26]$ and $[6,10]$ for
 set A and B respectively.}
 \label{fig:3pt}
% \vspace{-1.5ex}
\end{figure}

We simultaneously fit two sets of three-point functions to
eq.~(\ref{eq:3pt-2nd}) with $m_P$ fixed to the value extracted from the
two-point function.
As seen in Fig.~\ref{fig:3pt}, $t$-dependence of the three-point
functions are well described by eq.~(\ref{eq:3pt-2nd}).
Then the lattice $B$-parameter $B_P^{\rm lat}$ is obtained by
\begin{eqnarray}
     B_P^{\rm lat} 
 &=& \frac{3}{8}
     \left(\frac{2}{Z^{\rm wall}_{A_4}\,f_P}\right)^2
     \times
     \frac{(Z^{\rm wall}_{A_4})^2
           \langle \bar P|O^{\rm lat}_{L_\mu L_\mu}\ |P \rangle}
          {(2\,m_P)^2},
\end{eqnarray}
where the first and second factors are obtained from the two- and
three-point functions, respectively.
The fit range dependence of $B_P$ was studied, and found to be stable.

We adopt the RI-MOM scheme to calculate the renormalization factor.
Following the standard method, we obtain a preliminary result
\begin{eqnarray}
Z_{B_K}^{\rm RGI}=1.217(6),\ \ \
Z_{B_K}^{\overline{\rm MS}}(2\ {\rm GeV})=0.862(4) \ \ \
\mbox{(the error is statistical only)}.
\end{eqnarray}

\section{Test of NLO ChPT and extraction of $B_K$}

We first test whether the quark mass dependence of $B_P$ is consistent
with the NLO partially quenched ChPT (PQChPT) prediction, or to which
quark mass the prediction describes data well.
In the test, we only use data points which satisfy
$m_{\rm sea}\le m_{\rm valence}$ for the reason described below.

\begin{figure}
 \vspace{-0.5ex}
 \centering
  \includegraphics*[width=0.55 \textwidth,clip=true]
  {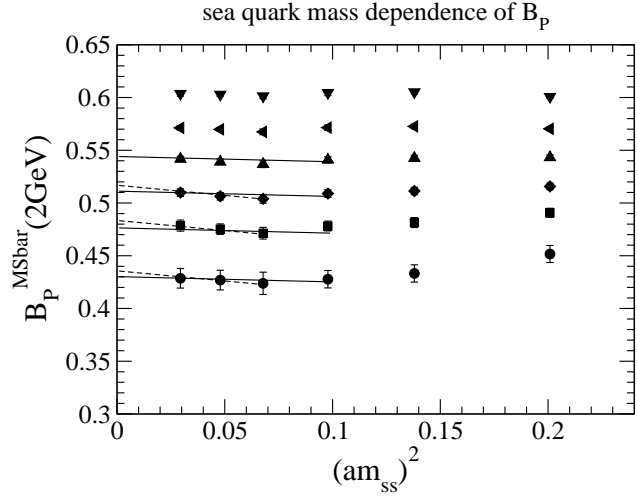}\\[-2ex]
 \caption{Sea quark mass dependence of $B_P^{\overline{MS}}$(2GeV).
 The different symbols denote different valence quark mass: 0.015--0.10
 from bottom to top.
 Only the data consisting of degenerate valence quarks are shown.
 Lines are the results from a linear fit and just a guide to eyes.}
 \label{fig:bp-mseadep}
% \vspace{-2ex}
\end{figure}

Figure~\ref{fig:bp-mseadep} shows the sea quark mass dependence of
$B_P$, in which clear dependence is not seen except for the region with
$m_{\rm sea}> m_{\rm valence}$.
In Ref.~\cite{Becirevic:2003wk}, the finite volume effects to $B_K$ were
studied to NLO in the framework of PQChPT, and found to become more
significant when $m_{\rm sea}> m_{\rm valence}$.
While the size effect found in Ref.~\cite{Becirevic:2003wk} is tiny, it
is pointed out in Ref.~\cite{Colangelo:2005gd} that the NLO estimate
significantly underestimates for $m_{\pi}$ and $f_\pi$.
For example, the NLO estimate of the size effect to $f_\pi$ gives about
2 \% correction at our lightest unquenched point while the inclusion of
NNLO gives 4--5 \%.
Motivated by these observations, we include the data point in the fit only
when $m_{\rm sea}\le m_{\rm valence}$.

\begin{figure}
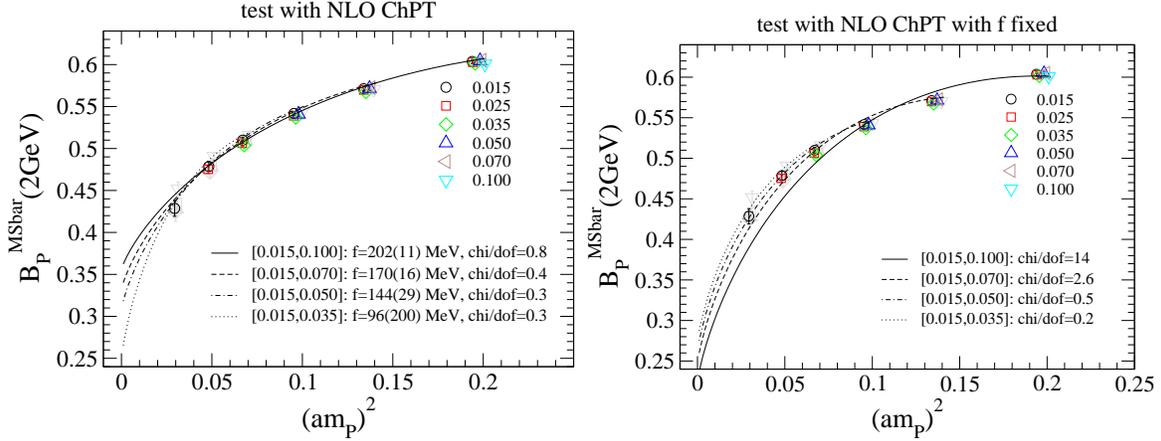

 \vspace{-1ex}
 \centering
  \includegraphics*[width=0.5 \textwidth,clip=true]
  {figs/bk-nlo-mv1=mv2_ffree-limit.eps}~
  \includegraphics*[width=0.5 \textwidth,clip=true]
  {figs/bk-nlo-mv1=mv2_ffix-limit.eps}\\[-1ex]
 \caption{Test with the NLO ChPT formula.
 The different symbols denote the different sea quark mass.
 The lines correspond to those in $m_{\rm sea}=0$.
 In the figure legend, [0.015,0.100] denotes the range of the sea and
  valence quark masses used in the fit, for example.}
 \label{fig:bp-valdep-fit2-limit}
 \vspace{-3ex}
\end{figure}

The test is made using data consisting of degenerate quarks.
$B_P$ is fitted to the NLO PQChPT
formula~\cite{Golterman:1997st,Becirevic:2003wk},
\begin{eqnarray}
     B_P
 &=& B_P^\chi\Bigg[\,
       1- \frac{6\,m_P^2}{(4\pi f)^2}
          \,\ln\left(\frac{m_P^2}{\mu^2}\right)\Bigg]
     + (b_1-b_3)\,m_P^2 + b_2\,m_{ss}^2,
 \label{NLOform}
\end{eqnarray}
where $m_{ss}^2\sim B_0(m_{\rm sea}+m_{\rm sea})$ and the free parameters
are $B^\chi_P$, $f$, $(b_1-b_3)$ and $b_2$.
The fit results are shown in Fig.~\ref{fig:bp-valdep-fit2-limit} (left).
$f$, the tree level pion decay constant with $f\sim$130 MeV
normalization, and $\chi^2/$dof are also shown for each fit range.
While all fit ranges tested give acceptable $\chi^2/$dof, $f$
monotonically decreases as the fit range is made narrower.
$f$'s obtained from the two narrowest ranges are consistent with each other
within one standard deviation, and its value is consistent with a naive
expectation $f=100\sim 130$ MeV.
Fixing $f$ to 110 MeV~\cite{Noaki:2007}, we obtain
Fig.~\ref{fig:bp-valdep-fit2-limit} (right).
The $\chi^2$/dof values suggest that while the PQChPT formula does not apply
for the two heaviest data points, the data for $m_q\le 0.050$ (roughly
corresponding to half strange mass) are inside the NLO ChPT regime.

To extract $B_K$, we fit the data of both degenerate and non-degenerate
quarks to  the following
formula~\cite{Golterman:1997st,Becirevic:2003wk,Aoki:2004ht},
\begin{eqnarray}
     B_{12}
 &=& B_{12}^\chi\Bigg[
       1- \frac{2}{(4\pi f)^2}
       \Bigg\{ m_{ss}^2 + m_{11}^2 -  \frac{3\,m_{12}^4+m_{11}^4}{2\,m_{12}^2}
        + m_{12}^2\left(     \ln\left(\frac{m_{12}^2}{\mu^2}\right)
                     + 2\,\ln\left(\frac{m_{22}^2}{\mu^2}\right)
               \right)
\nonumber\\&&\hspace*{18ex}
       - \frac{1}{2}\left( \frac{m_{ss}^2(m_{12}^2+m_{11}^2)}{2\,m_{12}^2}
                          + \frac{m_{11}^2(m_{ss}^2-m_{11}^2)}{m_{12}^2-m_{11}^2}
                     \right)\ln\left(\frac{m_{22}^2}{m_{11}^2}\right)
       \Bigg\}\Bigg]
\nonumber\\&&\hspace*{0ex}
     + b_1\,m_{12}^2
     + b_3\,m_{11}^2\left(-2+\frac{m_{11}^2}{m_{12}^2} \right)
     + b_2\,m_{ss}^2
     + c_1\,m_{11}^2\,m_{12}^2
     + \frac{c_2\,m_{12}^4}{1+c_3\,m_{12}^2+c_4\,m_{12}^4},
 \label{NLOform-nondege}
\end{eqnarray}
where $m_{ij}^2\sim B_0\,(m_{vi}+m_{vj})$ and $m_{vi}$ denotes a valence
quark mass.
%The above formula is valid when $m_{v1}<m_{v2}$.
%In the limit of $m_{v1}=m_{v2}$, the above formula without the last two
%terms reduces to eq.~(\ref{NLOform}).
The last two terms in eq.~(\ref{NLOform-nondege}) are added to describe
the data in the heavy region.
The fit is performed with four data sets, each set including
data from lightest three, four, five and six sea quarks.
\begin{figure}
 \vspace{-3ex}
 \centering
  \includegraphics*[width=0.6 \textwidth,clip=true]
  {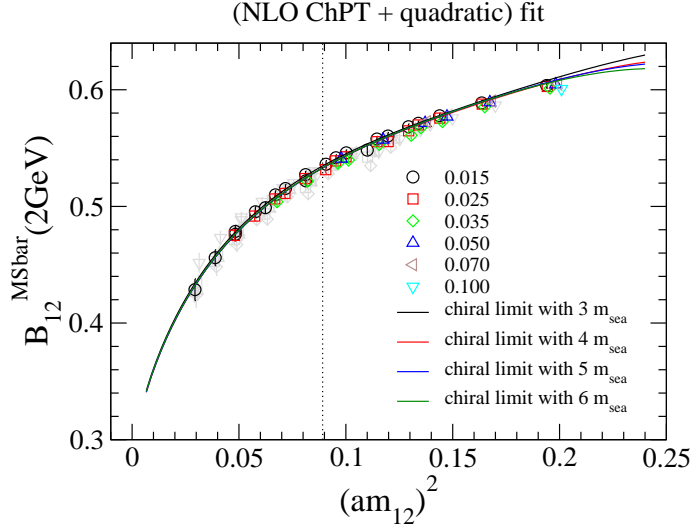}
 \vspace{-2ex}
 \caption{
 $m_{12}^2$ dependence of $B_{12}^{\overline{MS}}$(2GeV).
 The different symbols correspond to the different $m_{\rm sea}$.
 The solid lines represent $B_P$ extrapolated to
 $m_{\rm sea}=m_{v1}=m^{\rm phys}_{ud}$.
 The vertical line denotes the position of physical $m_K$.}
 \vspace{-1.5ex}
 \label{fig:bp-nnlo-nondege-limit}
\end{figure}

The fit results are shown in Fig.~\ref{fig:bp-nnlo-nondege-limit}.
The solid line is the one in which the lighter valence mass ($m_{v1}$)
and the sea quark mass ($m_{\rm sea}$) are extrapolated to the physical
$u, d$ mass ($m^{\rm phys}_{ud}$).
Interpolating to physical $m_K$, we obtain
$B_K^{\overline{\rm MS}}(2\ {\rm GeV})$=0.533--0.523 depending on the
data sets used.
As our preliminary result we take the result using four $m_{\rm sea}$
data, and obtain
\begin{eqnarray}
 B_K^{\overline{\rm MS}}(2\ {\rm GeV})=
%\left\{  \begin{array}{ll}
%   0.533(14) & \mbox{with 3 $m_{\rm sea}$ data}\\
   0.526( 9)  %\mbox{with 4 $m_{\rm sea}$ data}\\
%   0.523( 9) & \mbox{with 5 $m_{\rm sea}$ data}\\
%   0.525(26) & \mbox{with 6 $m_{\rm sea}$ data}
%  \end{array}
%    \right..
\end{eqnarray}
%\begin{eqnarray}
% B_K^{\overline{\rm MS}}(2\ {\rm GeV})=
%%  \left\{  \begin{array}{ll}
%%   0.535(7) & \mbox{with 3 $m_{\rm sea}$ data}\\
%   0.533(7) & \mbox{with 4 $m_{\rm sea}$ data}
%%    \\
%%   0.532(14) & \mbox{with 5 $m_{\rm sea}$ data}\\
%%   0.533(8) & \mbox{with 6 $m_{\rm sea}$ data}
%%  \end{array}    \right.,
% \label{eq:prelimi-final}
%\end{eqnarray}
where only the statistical error is shown.

Since in the above fit we did not include the data which could have
potentially significant finite size effect, the fit result
is expected to be under control.
As a conservative upper bound of the finite size effect, we take
that of $f_\pi$, and add a 5\% error.
%To have reliable estimation for this effects, simulations with different
%volume are necessary.

\section{The effect of fixing topology}

\begin{figure}
 \vspace{-2.5ex}
 \centering
  \includegraphics*[width=0.6 \textwidth,clip=true]
  {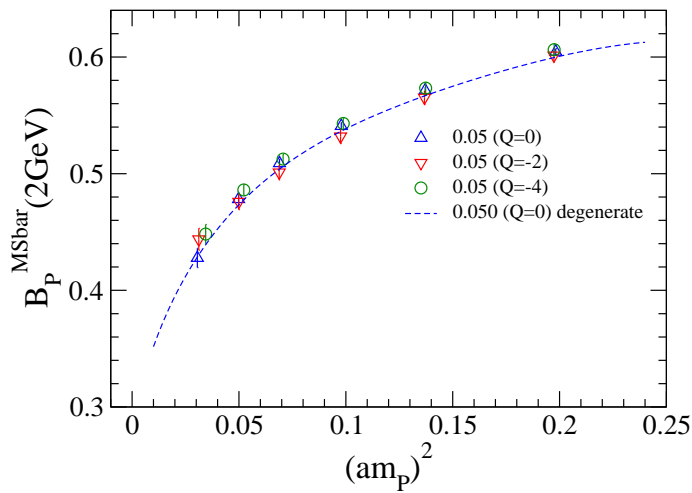}
 \vspace{-2ex}
 \caption{
 Comparison of $B_P^{\overline{MS}}$(2GeV) at $m_{\rm sea}$=0.05 with
 three different $Q$.
% The line represent is the result of the fit made in
% Dec.~X.X %\reff{subsec:usual fse}
% with $m_{v1}=m_{v2}$.
 }
 \label{fig:bp-nnlo-nondege-limit-topo}
 \vspace{-1.5ex}
\end{figure}

To estimate the effect of fixing the topological charge on $B_P$,
according to the studies in Refs.~\cite{Brower:2003yx,Aoki:2007ka},
we assume it to be
\begin{eqnarray}
 \sim \frac{m_{ps}^2}{(4\pi\,f)^2}\frac{1}{\langle Q^2 \rangle}
      \left(1-\frac{Q^2}{\langle Q^2 \rangle}\right),
\label{eq:assumption}
\end{eqnarray}
where $\langle Q^2 \rangle=\chi_t V_4\sim 10$ at
$m_q=0.05$~\cite{Ting-Wai:2007}.
This is motivated by an observation that the most significant
$\theta$-dependence of the physical quantities is that of pion mass, and
other quantities are affected through it.
Then, the correction to the $Q=0$ result is estimated to be 1.4\% at
$m_q$=0.05, and the difference between $Q=0$ and $-2$ $(-4)$ to be 0.6\%
(2.2\%).
Since the size of the statistical error for $B_P$ is about 2\%, one does
not expect to see clear $Q$ dependence of $B_P$.
In Fig.~\ref{fig:bp-nnlo-nondege-limit-topo}, $B_P$ at $m_{\rm sea}=0.05$
from three $Q$ are compared, where only the data of degenerate quarks are
shown.
We could not observe any systematic $Q$ dependence which is
statistically significant.
Thus the assumption eq.~(\ref{eq:assumption}) seems to give a reasonable
or even conservative estimate.
We will quote 1.4 \% as a crude estimate for the systematic error due
to fixing topology.
%Explicit estimation based on ChPT is necessary.

Since the calculation is made only at a single lattice spacing, it is
difficult to estimate systematic uncertainty due to scaling violation,
though this is expected to be under control as no $O(a)$ error is
present.
The renormalization factor $Z_{B_K}$ may have a sizable systematic error
as well.
The study to estimate all these errors is in progress.

\vspace{1.0ex}
Numerical simulations are performed on IBM System Blue Gene Solution at
High Energy Accelerator Research Organization (KEK) under a support of
its Large Scale Simulation Program (No. 07-16).
This work is supported in part by the Grant-in-Aid of the Ministry of
Education (No. 17740171, 18034011, 18340075, 18740167, 18840045,
19540286, 19740160).


\vspace{-0.5ex}
\begin{thebibliography}{99}

 \bibitem{Yao:2006px}
  W.~M.~Yao {\it et al.}  [Particle Data Group],
  %``Review of particle physics,''
  J.\ Phys.\ G {\bf 33}, 1 (2006).
  %%CITATION = JPHGB,G33,1;%%

 \bibitem{Fukaya:2006vs}
  H.~Fukaya, S.~Hashimoto, K.~I.~Ishikawa, T.~Kaneko, H.~Matsufuru,
  T.~Onogi and N.~Yamada [JLQCD Collaboration],
  %``Lattice gauge action suppressing near-zero modes of H(W),''
  Phys.\ Rev.\  D {\bf 74}, 094505 (2006)
  [arXiv:hep-lat/0607020].
  %%CITATION = PHRVA,D74,094505;%%

% \bibitem{Kaneko:2006pa}
%  T.~Kaneko {\it et al.}  [JLQCD Collaboration],
%  %``JLQCD's dynamical overlap project,''
%  PoS {\bf LAT2006}, 054 (2006)
%  [arXiv:hep-lat/0610036].
%  %%CITATION = POSCI,LAT2006,054;%%

 \bibitem{Becirevic:2003wk}
  D.~Becirevic and G.~Villadoro,
  %``Impact of the finite volume effects on the chiral behavior of f(K) and
  %B(K),''
  Phys.\ Rev.\  D {\bf 69}, 054010 (2004)
  [arXiv:hep-lat/0311028].
  %%CITATION = PHRVA,D69,054010;%%

 \bibitem{Colangelo:2005gd}
  G.~Colangelo, S.~Durr and C.~Haefeli,
  %``Finite volume effects for meson masses and decay constants,''
  Nucl.\ Phys.\  B {\bf 721}, 136 (2005)
  [arXiv:hep-lat/0503014].
  %%CITATION = NUPHA,B721,136;%%

 \bibitem{Golterman:1997st}
  M.~F.~L.~Golterman and K.~C.~L.~Leung,
  %``Applications of partially quenched chiral perturbation theory,''
  Phys.\ Rev.\  D {\bf 57}, 5703 (1998)
  [arXiv:hep-lat/9711033].
  %%CITATION = PHRVA,D57,5703;%%

 \bibitem{Noaki:2007}
  J.~Noaki {\it et al.} [JLQCD collaboration], in these proceedings.

 \bibitem{Aoki:2004ht}
  Y.~Aoki {\it et al.},
  %``Lattice QCD with two dynamical flavors of domain wall fermions,''
  Phys.\ Rev.\  D {\bf 72}, 114505 (2005)
  [arXiv:hep-lat/0411006].
  %%CITATION = PHRVA,D72,114505;%%

 \bibitem{Brower:2003yx}
  R.~Brower, S.~Chandrasekharan, J.~W.~Negele and U.~J.~Wiese,
  %``QCD at fixed topology,''
  Phys.\ Lett.\  B {\bf 560}, 64 (2003)
  [arXiv:hep-lat/0302005].
  %%CITATION = PHLTA,B560,64;%%

 \bibitem{Aoki:2007ka}
  S.~Aoki, H.~Fukaya, S.~Hashimoto and T.~Onogi,
  %``Finite volume QCD at fixed topological charge,''
  Phys.\ Rev.\ D {\bf 76}, 054508 (2007)
  [arXiv:0707.0396 [hep-lat]].
  %%CITATION = PHRVA,D76,054508;%%

 \bibitem{Ting-Wai:2007}
  T.~W.~Chiu {\it et al.} [JLQCD and TWQCD Collaboration], in these
  proceedings.
\end{thebibliography}
\end{document}